\documentstyle[aps,prl,epsf]{revtex}

\begin{document}

\newcommand\beq{\begin{equation}}
\newcommand\eeq{\end{equation}}
\newcommand\bea{\begin{eqnarray}}
\newcommand\eea{\end{eqnarray}}
\newcommand{\ket}[1]{| #1 \rangle}
\newcommand{\bra}[1]{\langle #1 |}
\newcommand{\braket}[2]{\langle #1 | #2 \rangle}
\newcommand{\proj}[1]{| #1\rangle\!\langle #1 |}
\newcommand{\ba}{\begin{array}}
\newcommand{\ea}{\end{array}}

\twocolumn[\hsize\textwidth\columnwidth\hsize\csname
@twocolumnfalse\endcsname

\author{Barbara M. Terhal, David P. DiVincenzo, and Debbie W. Leung}

\title{Hiding bits in Bell states}


\address{\vspace*{1.2ex}
            \hspace*{0.5ex}{IBM Watson Research Center,
P.O. Box 218, Yorktown Heights, NY 10598, USA}}

\date{\today}

\maketitle

\begin{abstract}
We present a scheme for hiding bits in Bell states that is secure
even when the sharers Alice and Bob are allowed to carry out local
quantum operations and classical communication. We prove that the information that Alice and Bob can gain about a hidden bit is exponentially small in $n$, the number of qubits in each share, and can be made arbitrarily small 
for hiding multiple bits. We indicate an alternative efficient low-entanglement method for preparing the shared quantum states. We discuss how our scheme
can be implemented using present-day quantum optics.
\end{abstract}


]

The protection of a secret by {\em sharing} it, that is, by
apportioning the secret data among two or more parties so that the
data only become intelligible as a consequence of their cooperative
action, is an important capability in modern information processing.
Here we give a method of using particular quantum states to share a
secret between two parties (Alice and Bob), in which the data is
hidden in a fundamentally stronger way than is possible in any
classical scheme.  We prove that even if Alice and Bob can communicate
via a classical channel, they can only obtain arbitrarily
little information about the hidden data.  They can unlock the secret
only by joint quantum measurements, which require either a quantum
channel, shared quantum entanglement, or direct interaction between
them.  We show that the creation of these secret shares can be done
with just a small expenditure of quantum entanglement: no more than one
Einstein-Podolsky-Rosen pair per secret bit shared. 

Our results are part of a larger exploration of the
information-theoretic capabilities of quantum mechanics, notable
examples of which (quantum key distribution \cite{bb84art} and quantum
teleportation \cite{tele}) have now begun to be realized in the
laboratory. The extent to which quantum states can hide shared data can be viewed as a new information-theoretic characterization of the quantum nonlocality of these states. Other workers have previously identified quantum secret sharing protocols \cite{cgl:secshar,got:secshar}, in
which participants (possibly more than two) receive shares of either
quantum or classical data.  In this previous work, however, there is
no guarantee that the data remains hidden if the parties choose to
communicate classically.  In fact, recent analysis
\cite{walgate:orthodist} has shown that, for a single hidden bit,
secrecy in the presence of classical communication is impossible if
the shares consist of parts of two orthogonal {\em pure} quantum
states.  This stronger form of data hiding is nonetheless possible,
as we will show, but only when the shares are made up from {\em mixed}
quantum states. 

Unlike usual secret sharing schemes, the security in our scheme does not depend on certain parties being honest or malevolent; we assume that both Alice and Bob are malevolent in the sense that they would go to any length to determine the hidden bit. The security of our scheme relies on the fact that 
Alice and Bob are restricted in their operations, a condition that could
be enforced by a third party. One can imagine, for example, a situation in which the third party (the boss) has a piece of data on which she would like Alice and Bob (some employees) to act without the sensitive data being revealed to them, or, in another scenario, the secret could be given to Alice and Bob and be revealed to them at a later stage determined by the boss.
Our scheme is such that at some later stage, the boss can
provide the employees with entanglement that enables the parties to
determine the secret with 100\% certainty. This last idea can in fact be used to establish a form of conditionally secure quantum bit commitment \cite{dlt:long}.
For these scenarios to work, we have to assume that the boss controls the 
(quantum) channel which connects the two parties: Alice and Bob 
are not allowed to communicate via a quantum channel. This prohibition can be enforced
by the boss by putting dephasing or noisy operations in their channel.
Furthermore, the boss controls the labs in which the employees operate; for example she can, prior to operation, sweep those labs clean of any entanglement (again by dephasing). 

We present our protocol and prove its security for a one-bit
secret $b$; at the end we indicate the proof of the security of its
multibit extension.  The protocol involves a ``hider'' (the boss above) who prepares one of two orthogonal bipartite quantum states
$\rho^{(n)}_{b=0}$ or $\rho^{(n)}_{b=1}$ based on the value of $b$,
and presents the two parts of the state to Alice and Bob.  $n$ is
an integer which determines the degree of security of the
protocol.  The hider is assumed to have a supply of each of
the four Bell states, defined as
$\ket{\Phi^{\pm}}=\frac{1}{\sqrt{2}}(\ket{00}\pm \ket{11})$ and
$\ket{\Psi^{\pm}}=\frac{1}{\sqrt{2}}(\ket{01}\pm \ket{10})$.
$\ket{\Psi^-}$ is a spin singlet while the other three are spin
triplets.  When $b=1$, the hider picks at random a set of $n$ Bell
states with uniform probability, except that the number of singlets
must be odd.  The $b=0$ protocol is the same, except that the number
of singlets must be even.  The hider distributes the $n$ Bell states
to Alice and Bob; for each Bell state the first qubit goes to Alice
and the second to Bob.  

 
To prove the security of this protocol, we must consider what
information Alice and Bob can gather about the bit $b$.  We assume
that Alice and Bob can perform any sequence of Local quantum Operations
supplemented by unlimited 2-way Classical Communication (we abbreviate this
class of operations as LOCC).  This class of operations does {\em not}
permit measurements in the basis of Bell states, from which the bit
could easily be determined: Alice and Bob simply count the number of
singlets measured and compute the parity.  In fact, we will show that
the information that Alice and Bob can learn about the hidden bit is
exponentially small in the number $n$ of Bell states that the hider
uses for the encoding.

Before analyzing the security of our protocol, let us pause and consider
the possibility for realizing our scheme in a physical experiment. The protocol that we have described above can be implemented in a present-day quantum optics lab in the following way. The hider needs to be able to 
make any one of the 4 Bell states; with an optical downconverter 
she can make a maximally entangled state between two polarization modes 
$\frac{1}{\sqrt{2}}(\ket{\updownarrow\,, \updownarrow}+\ket{\leftrightarrow\,,\leftrightarrow})$ \cite{mattleetal} and by further single qubit operations she can map this state onto any of the other three Bell states. The photons can be sent through two optical 
fibers to the Alice and Bob locations. Then Alice and Bob can attempt 
to unlock the secret by LOCC (in Ref. \cite{dlt:long} we will describe an
optimal and simple LOCC procedure, involving only 1-qubit gates).
For the complete unlocking of the secret, a quantum channel between Alice and Bob is opened up and Alice's photons are sent to Bob. Finally, Bob will need to do a measurement which distinguishes the singlet state $\frac{1}{\sqrt{2}}(\ket{\updownarrow\,,\leftrightarrow}-\ket{\leftrightarrow\,,\updownarrow})$ from the other three other Bell states. Such an incomplete measurement has been performed in the lab \cite{mattleetal}; a full Bell measurement is not necessary and is also not technologically feasible in current experiments.
Our alternative low-entanglement preparation scheme will be 
somewhat more involved in the lab but is interesting nonetheless. As we will discuss, what is needed are quantum operations in the Clifford group, including some particular one-qubit gates obtainable by linear optics in addition to the CNOT gate which cannot be implemented perfectly by using linear optical elements. However, 
recent work by Knill {\em et al.} \cite{klm} shows that a CNOT gate 
can be implemented near-deterministically in linear optics when single-photon sources are 
available.


Let us now pass to the security proof of our scheme. 
The LOCC class, even though it plays a fundamental role in the theory
of quantum entanglement, is remarkably hard to characterize succinctly
\cite{rainsdist}.  However, our analysis will rely on just one
important feature that all LOCC operations share: they cannot create
quantum entanglement between Alice and Bob if they are
initially unentangled.  We consider a general measurement scheme for
Alice and Bob that, irrespective of its precise physical
implementation, leads to just two final outcomes, ``0'' or ``1''.  
%
%
It can thus be described as a POVM (Positive Operator Valued
Measure) measurement~\cite{peresbook}, with two POVM elements, $M_0
\geq 0$ and $M_1\geq 0$, associated with outcomes ``0'' and ``1'' respectively.
%
%
In our case, $M_{0,1}$ operate on a Hilbert space of dimension $2^{2n}$, 
corresponding to the dimension of the input states.
For an input 
density matrix $\rho$, 
the outcome $b$ occurs with probability ${\rm Tr}(\rho M_b)$.  
Probability conservation implies that $M_0+M_1=I$, where $I$ is the identity matrix.
%
%

\begin{figure}[htb]
\begin{center}
\centerline{\epsfxsize=3in \epsfbox{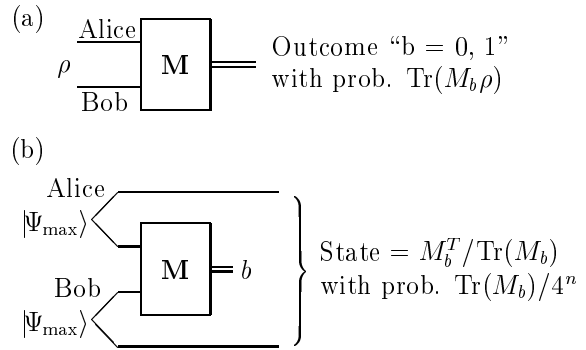}}
\vspace*{0.5cm}
\caption{
(a) A bipartite POVM measurement ${\bf M}$ performed on input $\rho$. Single horizontal lines are quantum registers, and double
lines are classical registers. The box represents a protocol (and circuit) for performing ${\bf M}$.
%
(b) A LOCC protocol that prepares a state proportional to $M_b^T$.
The two registers of the maximally entangled states $|\Psi_{\max}\rangle$ 
are represented by two lines connected in the far left.  The output 
probabilities are given by ${\rm Tr}\, \rho M_b$ with $\rho = I/4^n$.  
}
\label{fig1}
\end{center}
\end{figure}

A POVM measurement ${\bf M}$ for a bipartite input is depicted in 
Fig.~\ref{fig1}(a).
%
%
%
%
Such a POVM measurement implemented by LOCC cannot create quantum
entanglement.
This condition translates to two necessary conditions on the
measurement, $({\bf 1} \otimes T)[M_b] \geq 0$ for $b=0,1$.  Here, ${\bf 1}$ is the
identity operation on Alice's system, $T$ is the matrix transposition
on Bob's system, $T[\ket{i}\bra{j}]=\ket{j}\bra{i}$, and ${\bf 1}
\otimes T$ is called the {\em partial transpose} operation.
%
%
%
%
%
These conditions are proved as follows:    
Suppose Alice and Bob each prepares a maximally entangled state
$|\Psi_{max}\rangle = {1 \over \sqrt{2^n}} \sum_{i=0}^{2^n-1} |i\rangle |i\rangle$ in his own lab.
%
Then they apply the measurement ${\bf M}$, each on one register of $|\Psi_{\max}\rangle$, as illustrated in Fig.~\ref{fig1}(b).
When outcome $b$ is obtained, the residual state in the two 
unmeasured halves is proportional to
\bea
	\rho_f & \propto &
	\sum_{i,j,m,n} |i,j \rangle \langle m,n| ~ 
	{\rm Tr}\,[M_b \, |i,j \rangle \langle m,n|]
\nonumber
\\
	& = & \sum_{l,j,m,n} 
	\langle i,j| M_b^T |m,n \rangle |i,j \rangle \langle m,n| 
	= M_b^T \,,
\eea
where $M_b^T$ is the matrix transpose of $M_b$.  
%
Thus Fig.~\ref{fig1}(b) prescribes a LOCC procedure to create the states $M_b^T/{\rm
Tr}(M_b)$, since the input maximally entangled states are prepared by
Alice and Bob locally.
%
%
Therefore, the states $M_b^T/{\rm Tr}(M_b)$ are necessarily
unentangled and, following the Peres criterion \protect
\cite{Asher96}, they are positive under partial transposition (PPT),
meaning that $({\bf 1}\otimes T)[M_b^T]\geq 0$, which in turns implies 
$({\bf 1}\otimes T)[M_b] \geq 0$.


We now use the constraints that $M_{0,1}$ are PPT to bound the probability of a successful measurement. In particular we consider the probability $p_{0|0}$ that Alice and Bob decide for outcome ``0'' when the hider has prepared $\rho_0^{(n)}$ (corresponding to the hidden bit $b=0$), which is equal to $p_{0|0}={\rm Tr} M_0 \rho_0^{(n)}$. Similarly we define $p_{1|1}={\rm Tr}M_1 \rho_1^{(n)}$, the probability of outcome ``1'' when $\rho_1^{(n)}$ is prepared by the hider. 

First, we show that it is not necessary to consider the most general
pair of PPT operators $M_0$ and $M_1$. If there exists a general pair $(M_0,M_1)$ obeying
the PPT constraints, then there is another PPT pair
$(\tilde{M_0},\tilde{M_1})$ which is {\em diagonal} in the basis of
$n$ Bell states, such that the measurement with $\tilde{M_0}$ and
$\tilde{M_1}$ has the same probabilities of success $p_{0|0}$ and
$p_{1|1}$.
%
%
$\tilde{M_0}$ and $\tilde{M_1}$ are related to $M_0$ and $M_1$ by an
action called {\em partial twirling} \cite{bdsw} which removes all off-diagonal terms in the Bell basis and leaves the diagonal terms unchanged.

%
The argument involves three observations.  
(1) Partial twirling can be implemented by LOCC operations \cite{bdsw} which
preserve the PPT property \cite{pptnodist}.  Thus 
$({\bf 1}\otimes T)[\tilde{M}_{0,1}]\geq 0$.  
(2) The trace-preservation condition $M_0+M_1=I$ is invariant
under twirling, and therefore $\tilde{M_0}+\tilde{M_1}=I$.  
%
%
(3) The states to be measured, $\rho_0^{(n)}$ and $\rho_1^{(n)}$, are
mixtures of tensor products of $n$ Bell states and thus are
Bell-diagonal. It follows that $p_{0|0}={\rm Tr} M_0 \rho_0^{(n)} = {\rm Tr}
\tilde{M_0} \rho_0^{(n)}$ and likewise $p_{1|1} = {\rm Tr}\tilde{M_1}
\rho_1^{(n)}$, because the off-diagonal terms of $M_{0,1}$ do not 
contribute to the trace. This establishes the argument; we can, without loss of generality, restrict to a measurement with Bell diagonal POVM elements.


To carry the analysis further we introduce a compact
notation\cite{bdsw} that represents each of the four Bell states by two bits, as follows: $\ket{\Phi^+} \rightarrow 00$, $\ket{\Phi^-}
\rightarrow 01$, $\ket{\Psi^+} \rightarrow 10$ and the singlet
$\ket{\Psi^-} \rightarrow 11$.  A product of $n$ Bell states is thus
represented by a $2n$-bit string ${\bf s}$. The four Bell states can be
rotated into each other by local Pauli-matrix rotations, involving one
half of the entangled state only.  In the language of binary strings, we
can also associate two bits with each of the Pauli matrices, $\sigma_x
\rightarrow 10$, $\sigma_z \rightarrow 01$, $\sigma_y \rightarrow 11$
and $I \rightarrow 00$.  This notation is convenient because the
Pauli matrices then act on the two bits characterizing the Bell state
by a bitwise XOR. For example $(\sigma_z \otimes I)\ket{\Phi^+}=\ket{\Phi^-}$ can be represented as $01\oplus 00=01$. Using the identity 
\beq
\ba{l}
({\bf 1}\otimes T)[\ket{\Phi^+}\bra{\Phi^+}]= \\
\frac{1}{2}(\ket{\Phi^+}\bra{\Phi^+}+\ket{\Phi^-}\bra{\Phi^-}+
\ket{\Psi^+}\bra{\Psi^+}-\ket{\Psi^-}\bra{\Psi^-})
\ea
\eeq
permits the operators $({\bf 1} \otimes T)^{\otimes n}[M_{0,1}]$ to be written
very compactly in the binary-string notation.  We denote the diagonal
matrix elements of $M_0$ and $M_1$ in the basis of products of $n$
Bell states (labeled by the $2n$-bit string ${\bf s}$) by $\alpha_{\bf
s}$ and $\beta_{\bf s}$ respectively.  Using the fact that strings of
Bell states can be converted to each other by local Pauli operations,
we can compute the diagonal matrix elements of the equation $({\bf 1}
\otimes T)^{\otimes n}[M_0]\geq 0$ in the binary-string notation. We obtain 
the condition 
\beq
\sum_{{\bf s}} \alpha_{{\bf s} \oplus {\bf m}}(-1)^{N_{11}({\bf s})}
\geq 0\,,
\eeq
for all $2n$-bit strings ${\bf m}$, where $N_{11}({\bf s})$ is
the number of times that a $11$ pair appears in the binary string {\bf s}.
Through the association of Bell states with $2n$-bit strings,
$N_{11}({\bf s})$ is precisely the number of singlets $\ket{\Psi^-}$
among the set of $n$ Bell states.  The same calculation for $M_1$
gives $\sum_{{\bf s}} \beta_{{\bf s} \oplus {\bf m}} (-1)^{N_{11}({\bf
s})} \geq 0$ for all ${\bf m}$.  With the relation $\alpha_{\bf
s}=1-\beta_{\bf s}$, resulting directly from $M_0+M_1=I$, and
the identity $\sum_{\bf s}(-1)^{N_{11}({\bf s})}=2^n$ (which can shown
by evaluating a simple binomial sum), we obtain that, for all
$2n$-bit strings ${\bf m}$
\begin{equation}
0 \leq \sum_{{\bf s}} \alpha_{{\bf s} \oplus {\bf m}}
(-1)^{N_{11}({\bf s})} \leq 2^n.
\end{equation}
By setting ${\bf m}={00 \ldots 00}$ in this equation, we can express
the probabilities of success, $p_{0|0}=2/(2^{2n}+2^n)\sum_{{\bf
s}|N_{11}({\bf s})\ \mbox{\tiny is even}}\alpha_{\bf s}$ and
$p_{1|1}=2/(2^{2n}-2^n)\sum_{{\bf s}|N_{11}({\bf s})\ \mbox{\tiny is
odd}}\beta_{\bf s}$, in terms of these two inequalities.  This result
bounds the sum $p_{0|0}+p_{1|1}-1$ in both ways
\begin{equation}  
-\delta\leq p_{0|0}+p_{1|1}-1 \leq\delta\,,
\end{equation}
where $\delta=1/2^{n-1}$.  This result establishes the hiding
property: for $\delta=0$ (corresponding to $n\rightarrow\infty$),
Alice and Bob's measurement outcomes can be faithfully simulated by a
coin flip with bias $p_{0|0}$, and so give no information about the
identity of the state.  There is also an information-theoretic
interpretation of this result; we can show\cite{dlt:long} that, as a
consequence of these inequalities, the mutual
information\cite{cover&thomas:infoth} $I(B:M)$ is bounded by $\delta H(B)$,
where $B$ is the bit value and $M$ is the outcome of {\em any} LOCC
measurement by Alice and Bob, not just a two-outcome one. Here $H(B)$ \cite{cover&thomas:infoth} is 
the Shannon information of the hidden bit, which equals 1 in the case of 
equal prior probabilities for $b=0$ and $b=1$.


We now return to the question of how the hider can produce the states
$\rho_0^{(n)}$ and $\rho_1^{(n)}$ using minimal entanglement between
the two shares. We will demand that the procedure to create $\rho_0^{(n)}$ and $\rho_1^{(n)}$ is efficient as a quantum computation, that is, since each 
hiding state consists of $2n$ qubits, we seek a procedure to create these
states with little entanglement, using a number of quantum computation steps polynomial in $n$. 


We use a convenient alternative representation of these two density matrices: 
\bea
\rho_0^{(n)}=q_n \rho_1^{(n-1)}\otimes \rho_1^{(1)}+(1-q_n) \rho_0^{(n-1)} \otimes \rho_0^{(1)}, \nonumber \\
\rho_1^{(n)}=p_n \rho_0^{(n-1)}
\otimes \rho_1^{(1)}+(1-p_n)\rho_1^{(n-1)} \otimes
\rho_0^{(1)}. 
\label{recur1}
\eea
The mixing coefficients are
determined by the uniformity of the Bell mixtures and proper normalization:
\beq
\ba{lr}
q_n=\frac{2^{n-1}-1}{2(2^n+1)}, &\;\; p_n=\frac{2^{n-1}+1}{2(2^n-1)}.
\ea
\eeq
This representation can be easily understood by realizing that in 
order to make, say, a mixture of $n$ Bell states with an even number of singlets,
we can take a mixture of $n-1$ Bell states with an odd number of singlets 
and an additional singlet {\em or} (with the appropriate probability) a mixture of $n-1$ Bell states with
an even number of singlets and another Bell state which is not a singlet.

Solving the recurrence relations for these two density matrices, we
find that $\rho_0^{(n)}$ and $\rho_1^{(n)}$ are both so-called Werner
density matrices \cite{werner:lhv}: linear combinations of the identity matrix $I$ and the matrix $H=({\bf
1}\otimes T)^{\otimes n}[\ket{\Phi^+}\bra{\Phi^+}^{\otimes n}]$. In
particular, $\rho_0^{(n)} \propto I+2^n H$ and
$\rho_1^{(n)} \propto I-2^n H$.  It is known from previous
work that the Werner state $\rho_0^{(n)}$ is unentangled
\cite{nptnond1}.  In fact, we can show\cite{dlt:long} that it is possible
to make $\rho_0^{(n)}$ by first choosing a random element $U$ of the
Clifford group $C_n$\cite{gotthesis} and then applying
$U\otimes U$ on the state $\ket{0}^{\otimes n}\otimes \ket{0}^{\otimes n}$, i.e. the hider applies
the same rotation $U$ on both $n$-qubit shares of the state.  It can
be shown that this procedure takes $O(n^2)$ 1-qubit and 2-qubit
gates \cite{gotthesis} and polynomial classical computation
\cite{dlt:long}.  On the other hand, the Werner state $\rho_1^{(n)}$
is entangled; its entanglement of formation is known to be one ebit
\cite{vollwerner}.  Using Eq. (\ref{recur1}) for
$\rho_1^{(n)}$ and the fact that $\rho_0^{(n)}$ is unentangled, we
show explicitly how the hider can recursively create $\rho_1^{(n)}$ using just
one singlet: (1) The
hider flips a coin with bias $p_n$ for 0, and bias $1-p_n$ for 1. (2)
If the outcome is 0, then the hider prepares a tensor product of
$\rho_0^{(n-1)}$ and one singlet $\ket{\Psi^-}\bra{\Psi^-}$. This
costs one ebit, since $\rho_0^{(n-1)}$ is unentangled. If the outcome
is 1, then she prepares $\rho_1^{(n-1)} \otimes \rho_0^{(1)}$. Here
$\rho_0^{(1)}$ again requires no entanglement, and $\rho_1^{(n-1)}$
can similarly be prepared by the process just described.

Finally, we note that the obvious extension of the protocol
presented here permits the sharing of an arbitrary number of bits.
The hider simply encodes every bit in a different block of Bell states
as discussed above. The security analysis is more involved, since 
it cannot be excluded that joint measurements on all tensor 
product components provide more information than a measurement on each component separately. By exploiting the symmetry of the hiding states 
as expressed by their representation as Werner states, we are able to 
bound the mutual information $I({\bf B}:M)=I(B_1 B_2 \ldots B_k: M)\leq \epsilon$, where $M$ is now any multi-state random variable obtained from a measurement scheme on the $k$ encoded bits, provided that $n$, the number of Bell states in each block encoding $B_i$, scales as $n(k)\rightarrow 2k+\log k +\log \log e+\log 1/\epsilon$ in the large $k$ limit. This result has been derived \cite{dlt:long} for the
case of equal prior probabilities $1/2^k$ for all $k$-bit strings.


In conclusion, we have shown how to share bits in a pair of quantum
states such that an Alice and a Bob who do not share quantum
entanglement and cannot communicate quantum data, can learn
arbitrarily little information about the bits, whereas Alice and Bob
can obtain the bits reliably if they are given these resources.  



We would like to thank Charles Bennett for
sharing his insights in the problem of hiding bits with Bell states
and giving valuable comments on this paper. We would like to thank
John Smolin for interesting discussions and perspectives on the
results. We acknowledge support of the National Security Agency and
the Advanced Research and Development Activity through Army Research
Office contract number DAAG-55-98-C-0041.

\bibliographystyle{h-physrev}
\bibliography{refs}

\begin{thebibliography}{10}

\bibitem{bb84art}
C.~H. Bennett and G.~Brassard,
\newblock {\em Proc. of the IEEE Int. Conf. on Computers, Systems and Signal
  Processing} , 175 (1984).

\bibitem{tele}
C.~Bennett {\em et~al.},
\newblock Phys. Rev. Lett. {\bf 70}, 1895 (1993).

\bibitem{cgl:secshar}
R.~Cleve, D.~Gottesman, and H.-K. Lo,
\newblock Phys. Rev. Lett. {\bf 83}, 648 (1999), quant-ph/9901025.

\bibitem{got:secshar}
D.~Gottesman,
\newblock Phys. Rev. A {\bf 61}, 042311 (2000), quant-ph/9910067.

\bibitem{walgate:orthodist}
J.~Walgate, A.~Short, L.~Hardy, and V.~Vedral,
\newblock Phys. Rev. Lett. {\bf 85}, 4972 (2000), quant-ph/0007098.

\bibitem{dlt:long}
D.~DiVincenzo, D.~Leung, and B.~Terhal,
\newblock Quantum data hiding,
\newblock in preparation.

\bibitem{mattleetal}
K.~Mattle, H.~Weinfurter, P.~Kwiat, and A.~Zeilinger,
\newblock Phys. Rev. Lett. {\bf 76}, 4656 (1996).

\bibitem{klm}
E.~Knill, R.~Laflamme, and G.~Milburn,
\newblock Nature {\bf 409}, 46 (2001).

\bibitem{rainsdist}
E.~M. Rains,
\newblock Phys. Rev. A {\bf 60}, 173 (1999).

\bibitem{peresbook}
A.~Peres,
\newblock {\em Quantum Theory: Concepts and Methods} (Kluwer Academic
  Publishers, 1993).

\bibitem{Asher96}
A.~Peres,
\newblock Phys. Rev. Lett. {\bf 77}, 1413 (1996).

\bibitem{bdsw}
C.~Bennett, D.~DiVincenzo, J.~Smolin, and W.~Wootters,
\newblock Phys. Rev. A {\bf 54}, 3824 (1996), \mbox{arXive} eprint
  quant-ph/9604024.

\bibitem{pptnodist}
M.~Horodecki, P.~Horodecki, and R.~Horodecki,
\newblock Phys. Rev. Lett. {\bf 80}, 5239 (1998), quant-ph/9801069.

\bibitem{cover&thomas:infoth}
T.~M. Cover and J.~A. Thomas,
\newblock {\em Elements of Information Theory} (Wiley, 1991).

\bibitem{werner:lhv}
R.~Werner,
\newblock Phys. Rev. A {\bf 40}, 4277 (1989).

\bibitem{nptnond1}
D.~DiVincenzo, P.~Shor, J.~Smolin, B.~Terhal, and A.~Thapliyal,
\newblock Phys. Rev. A {\bf 61}, 062312 (2000), quant-ph/9910026.

\bibitem{gotthesis}
D.~Gottesman,
\newblock {\em Stabilizer Codes and Quantum Error Correction},
\newblock PhD thesis, CalTech, 1997, quant-ph/9705052.

\bibitem{vollwerner}
K.~Vollbrecht and R.~Werner,
\newblock Entanglement measures under symmetry,
\newblock arXive e-print quant-ph/0010095.

\end{thebibliography}

\end{document}